\documentstyle[epsf]{article}
\begin{document} 
\def\1{{\rm 1 \kern -.10cm I \kern .14cm}} \def\R{{\rm R \kern -.28cm I 
\kern .19cm}}

\begin{titlepage} 
\begin{flushright}  LPTHE-ORSAY 98/41 \\UFIFT-HEP-98-15 \\
CERN-TH/98-213 \\ hep-th/9807079 \\ 10 July 1998 \end{flushright} 
\vskip .8cm 
\centerline{\LARGE{\bf {Pseudo-anomalous $U(1)$ symmetry in the
strong}}}  \vskip .5cm
\centerline{\LARGE{\bf {coupling limit of the
heterotic string }}}    \vskip 1.5cm
\centerline{\bf {Pierre Bin\'etruy$^{a,b}$, C\'edric Deffayet$^a$,
Emilian Dudas$^{b,a}$  and Pierre Ramond$^c$\footnote{Supported in part by 
the 
United States Department of Energy under grant DE-FG05-86-ER40272.}}}   
\vskip
.5cm \centerline{$^a${\em Laboratoire de Physique Th\'eorique et
Hautes  Energies\footnote{Laboratoire associ\'e au CNRS-URA-D0063.}}}
\centerline{\em Universit\'e Paris-Sud, B\^at. 210,}
\centerline{\em F-91405 Orsay Cedex, France }
\vskip .5cm
\centerline{$^b${\em CERN, Theory Division,}}
\centerline{\em 1211 Geneva 23, Switzerland} 
\vskip .5cm
\centerline{$^c${\em Institute for Fundamental Theory,}}
\centerline{\em Department of Physics, University of Florida}
\centerline{\em Gainesville FL 32611, USA}

\vskip 1cm
\centerline{\bf {Abstract}}
We discuss, in the context of the strongly-coupled $E_8 \times E_8$
heterotic string proposed by Ho\u{r}ava and Witten, the appeareance of 
anomalous $U(1)_X$ symmetries of a nonperturbative origin, related to the
presence, after compactification, of five-branes in the five-dimensional
bulk of the theory. We compute the gauge anomalies and the induced
Fayet-Iliopoulos terms on each boundary, which we find to be lower
than the universal one induced in the weakly coupled case.
  
\indent

\end{titlepage}



The role played by a pseudo-anomalous abelian symmetry
\cite{DSW} present in many models constructed from the weakly coupled
heterotic string theory has been increasingly recognized. It may hold
a function as a family symmetry to explain quark and lepton mass
hierarchies \cite{IbRo}-\cite{FaPa}, as a mediator of supersymmetry
breaking  \cite{BiDu}-\cite{ADM}, in cosmology
\cite{CMMQ,Dterm}. The anomaly cancellation mechanism is a
four-dimensional remnant of the Green-Schwarz mechanism of anomaly
cancellation which makes use of the coupling of the dilaton-axion to
the gauge degrees of freedom. The physics of this anomalous
$U(1)$ depends primarily on the scale $\xi$ at which the
corresponding symmetry is broken. This scale may be computed from the
underlying string theory and lies one or two orders of magnitude
below the Planck scale, which may be suitable for family symmetry
purposes but probably too high for cosmology \cite{cosmo,BDP}.

In this article we discuss the possible origin of an anomalous $U(1)$
symmetry in the context of the strong coupling limit of the heterotic
string constructed by Ho\v{r}ava and Witten \cite{HoWi}. In this
picture, the observable and hidden gauge degrees of freedom live on
two distinct 10-dimensional boundary planes. Because of its
non-vanishing mixed anomalies, the anomalous $U(1)$ couples to both
types of degrees of freedom and must therefore be found in the
11-dimensional bulk. This restricts its possible origin and makes the
eleventh orbifold-like dimension between the two boundary planes play a
key role in unravelling its structure. This may also help, as we will
see, to evaluate the associated scale $\xi$.


Let us start by recalling some of the properties of the
pseudo-anomalous $U(1)_X$ as it appears in the weakly coupled
heterotic string.  The anomaly cancellation rests on the universal
coupling of the dilaton superfield $S$ to the gauge degrees of freedom:
\begin{equation}
{\cal L} = -{1 \over 4} {\rm Re} \; S \sum_a k_a F^{a\mu \nu}
F^a_{\mu \nu} + {1 \over 4} {\rm Im} \; S \sum_a k_a F^{a\mu \nu}
\tilde F^a_{\mu \nu},
\end{equation}
where $k_a$ is the Kac-Moody level of the gauge symmetry group $G_a$.
Anomaly cancellation is thus ensured by a Peccei-Quinn transformation
of the axion string Im $S$ at the sole condition that the mixed 
$U(1)_X-G_a-G_a$ anomaly coefficients $C_a$ satisfy:
\begin{equation}
{C_1 \over k_1} = {C_2 \over k_2} = \cdots = {C_a \over k_a} \equiv 
\delta_{GS}.
\end{equation}
The Green-Schwarz coefficient $\delta_{GS}$ is  non-vanishing  
which imposes that the mixed anomaly
coefficients for both observable and hidden sector gauge symmetries
are non-zero. This in turn implies that fermions of both sectors
are charged under the anomalous symmetry which therefore couples
observable and hidden sector. This will be a key property to help us
identify the origin of similar anomalous symmetries in the context of
the strongly coupling limit of the heterotic string.


In the Ho\v{r}ava-Witten construction, this limit is described by
eleven-dimensio\-nal supergravity compactified on ${\bf R}^{10} \times
{\bf S}^1 / Z_2$, coupled with gauge fields which are ten-dimensional
vector multiplets that propagate on the boundary of spacetime.
The corresponding bosonic action reads:
\begin{eqnarray}
{\cal S} &=& {1 \over \kappa_{11}^2} \int d^{11} x
\sqrt{g^{(11)}} \left( -{1 \over 2} {\cal R}^{(11)}
- {1 \over 48} G_{IJKL} G^{IJKL} \right) \nonumber \\
& & -{ \sqrt{2} \over 3456 \kappa_{11}^2} \int d^{11} x
\epsilon^{I_1 \cdots I_{11}} C_{I_1I_2I_3} G_{I_4 \cdots I_7}
 G_{I_8 \cdots I_{11}} \nonumber \\
& &-{1 \over 8 \pi (4 \pi \kappa_{11}^2)^{2/3}}
\int d^{10} x \sqrt{g^{(10)}} \; {\rm tr} F_{AB} F^{AB}, \label{eq:fullLag}
\end{eqnarray}
where $I,J,\cdots$ are eleven-dimensional indices, $A,B,\cdots$
ten-dimensional indices and $G$ is the field strength of the
three-form $C$ ($G=6dC+ \cdots$). The $Z_2$ projection corresponds to
the reflection on the eleventh coordinate ($x^{11} \rightarrow
-x^{11}$) and acts as the chirality projector on the gravitino degrees
of freedom. The 3-form $C$ is odd under this projection whereas the
metric tensor is even.

The bosonic action (\ref{eq:fullLag}) and the corresponding fermionic
one constructed in \cite{HoWi} do not form the complete quantum action
but they are rather an effective description including the lowest orders of an
expansion in the parameter $\kappa_{11}^{2/3}$. Because of the presence
of the boundary, the fermionic action includes divergent terms
proportional to $\delta(0)$ (as well as its derivatives) possibly to
some power: in a full quantum treatment, the boundary presumably
acquires a non-zero thickness of order $M^{-1}$ and the divergent
$\delta(0)$ terms are smoothed out into terms of order $M$, where  
$M$ is the  fundamental mass scale:
\begin{equation}
4 \pi \kappa^2_{11} = (2 \pi M^{-1})^9.
\end{equation}

One may compactify this theory down to 5 dimensions 
\cite{BaDi}-\cite{ELPP}. With a standard
embedding of the $SU(3)$ holonomy group of the 6-dimensional 
compact manifold,
one may consider that the ($E_6$-type) gauge fields of the observable
sector live on one boundary, whereas the ($E_8$-type) gauge fields of
the hidden sector live on the other one. The different scales involved
will play an important part in what follows. Let us therefore review
them. We will adopt a simplified  compactification scheme
\cite{Witten}  which includes the most generic
properties of more realistic scenarios: we keep only the two moduli
which describe respectively the radii of the six-dimensional compact
manifold (compactification from 11 to 5 dimensions) and of the
orbifoldlike 11th dimension (counted from now on as the fifth
dimension). In terms of the eleven-dimensional metric $g^{(11)}_{IJ}$
one may write:
\begin{eqnarray}
g^{(11)}_{ab} = e^\sigma g^{(0)}_{ab}, & \; \; \int d^6x
\sqrt{g^{(0)}} = (2 \pi M^{-1})^6 \nonumber \\
g^{(11)}_{\mu \nu} = e^{a \sigma} g^{(5)}_{\mu \nu} &
\label{eq:rescale1} \end{eqnarray}
where $a,b \in \{ 6,\cdots , 11\}$ and $\mu, \nu \in \{ 1, \cdots ,
5\}$, and
\begin{eqnarray}
g^{(5)}_{55} = e^{2 \gamma} \hat g^{(0)}, & \; \; \int dx^5
\sqrt{\hat g^{(0)}} = \pi M^{-1} \nonumber \\
g^{(5)}_{mn} = e^{b \gamma} g_{mn} &
\label{eq:rescale2} \end{eqnarray}
where $m,n \in \{ 1, \cdots ,4\}$. In these formulas, $M$ is the 
fundamental mass scale but the rescalings
undergone by the four-dimensional metric may change its
physical interpretation in 4-dimensional spacetime. In the absence
of rescaling ($a=b=0$) it is simply the fundamental scale of the
original eleven-dimensional theory, usually\footnote{ In fact, our
$M_{11}$ is $\ell^{-1}_{11}$ of ref. \cite{CKM} and $2\pi M_{11} 
(4\pi)^{-1/9}$ of ref. \cite{HoWi}.} denoted
by $M_{11}$. A look at the Einstein term in the four-dimensional
Lagrangian shows that the choice $a=-2,b=-1$ yields the
4-dimensional Planck scale $M=m_{Pl}\sqrt{2\pi}$ ($m_{Pl}$ being the
reduced Planck scale). Finally, the choice
$a=-2,b=0$ corresponds to the 5-dimensional Planck scale \cite{DuGr} as well as
5-brane unit mass scale in the 10-dimensional theory
\cite{5brane}; we will denote it by $M=M_5$. 

Our simplified compactification scheme amounts to the presence of
only the dilaton $S$ and a single K\"ahler modulus $T$. Their real
parts $s$ and $t$ expressed in general $M$ units simply read:
\begin{equation}
s = e^{3 \sigma}, \; \; \; t=e^\gamma e^{(a+2) \sigma/2}.
\end{equation}
The mass scale which corresponds to the inverse radius of the
6-dimensional compact manifold is the scale where the theory becomes
11-dimensional and therefore corresponds to the unification of
all couplings; we denote it by $M_U$. In original units, it simply
reads:
\begin{equation}
M_U = {M_{11} \over s^{1/6}}.
\end{equation}
One can easily obtain from (\ref{eq:fullLag}) the gauge kinetic terms in
four dimensions:
\begin{equation}
{\cal L} = -{1\over 4} \int d^4 x \sqrt{g} {s \over 2 \pi} F^{mn}
F_{mn}, \label{per0}
\end{equation}
which shows that the gauge coupling at the unification
scale $\alpha_U=g^2_U/(4\pi)$ simply reads:
\begin{equation}
\alpha_U = {1 \over 2s} \ . \label{per}
\end{equation}
Notice that this gauge coupling is universal at the tree level, for all
the gauge groups living on the boundary, which leads to the universality
of the Green-Schwarz mechanism in four dimensions.

In what follows, we will be interested mainly in expressing the scales 
in 5-dimensional units. One easily obtains the following relations:
\begin{equation}
m_{Pl}=M_5 \left( {t\over 2 \pi} \right)^{1/2}, \; M_U = M_5 s^{-1/2}, 
\; R_5^{-1} = M_5 t^{-1},
\end{equation}
where $R_5$ is the radius of the orbifold dimension. One then obtains:
\begin{eqnarray}
M_5 &=& M_U  (2 \alpha_U)^{-1/2} \sim 3.4 \; M_U \ , \nonumber \\
R_5^{-1} &=& {1 \over 2 \pi} {M_U^3 \over m_{Pl}^2} (2 \alpha_U)^{-3/2}
\sim 10^{-3} \; M_U \ , \label{eq:data} \\
M_{11} &=& M_U (2 \alpha_U)^{-1/6} \sim 1.5 \; M_U \ , \nonumber
\end{eqnarray}
where we have used $M_U = 3 \; 10^{16}$ GeV, $\alpha_U^{-1} = 23.3$ and
$m_{Pl} = M_{Pl}/\sqrt{8\pi} = 2.4 \; 10^{18}$ GeV.

It is important to note that, because the original theory is not
fully determined by the action (\ref{eq:fullLag}), its compactified
version is valid only for a certain range of mass scales. In particular,
we disregarded the non-zero thickness of the boundary, presumably
associated with some non-perturbative effect in quantum M-theory. Had we
restored a non-vanishing thickness, the gauge fields of the boundary
would propagate in the corresponding layer (of width of order $M^{-1}$):
this would generate in the 4-dimensional theory massive states of mass
$M$. Since we consider on the other hand the Kaluza-Klein states of mass
$R_5^{-1}$, our treatment is not consistent unless we impose the
condition (in our 5-dimensional units, $M=M_5$)\footnote{ This may also
be seen in a more technical way: the delta functions 
$\delta $ which appear are
invariant and therefore incorporate a factor $1/\sqrt{g_{5,5}}$ which is
$1/t$ in our 5-dimensional units; thus expansion in the
number of $\delta$ factors amounts in 4 dimensions to an expansion 
in $t^{-1}$ in our units where the rescaling of the 4-dimensional metric
is $t$-independent ($b=0$ in (\ref{eq:rescale2})). Similarly derivatives
such as $\delta'$ include a factor $1/g_{5,5}$ and therefore yield
higher powers in $1/t$.}:
\begin{equation}
M_5 R_5 = t \gg 1 \label{p0}.
\end{equation}
This is obviously verified if we plug in the data
(\ref{eq:data}). Notice that, physically, $M_5R_5$ is the number of 
Kaluza-Klein states of mass less than $M_5$, which contributes to computations
involving Kaluza-Klein states running in loops. 


Let us now turn to the anomalous $U(1)$ symmetry in this context. Since
it necessarily couples the observable and the hidden sector which lie on
the two boundaries of 11-dimensional spacetime, the corresponding gauge
degrees of freedom necessarily live in the 11-dimensional bulk (or at
least, when we consider the compactified theory, in the 5-dimensional
bulk). Indeed, consider the limiting case when the two boundaries are far apart,
$t \rightarrow \infty$ and therefore they do not interact with each other.
In this case, the $U(1)_X$ gauge coupling should vanish and therefore the $U(1)_X$
gauge group is intimately related to the presence of the extra dimension. In particular,
the heterotic perturbative gauge group, with a gauge coupling given by (\ref{per})
does not satisfy this constraint and cannot describe, in the M-theory regime, a
perturbative physics.  
  
An obvious candidate for our anomalous $U(1)_X$ would be the 5-dimensional gauge field 
$C_{\mu IJ}$ (among which is found the graviphoton which we denote by
$C_\mu$) but it is odd under the Ho\u{r}ava-Witten $Z_2$ parity and
therefore only the 4-dimensional scalar field $C_{5IJ}$ has
non-vanishing zero modes on the boundaries. We thus have to
assume that an anomalous $U(1)_X$ symmetry has a non-perturbative
origin (from the point of view of the weakly coupled heterotic string)
 and that the corresponding gauge field is even under the $Z_2$ 
parity. This is indeed a generic situation after compactification, where
the 5d bulk contain nonperturbative gauge fields and charged matter coming from
5-branes. The perturbative gauge fields on the boundary are
interpreted in an open string language as coming from 9-branes. There
are also mixed, $5-9$ sectors, corresponding, in an effective
Horava-Witten type lagrangian, to boundary fields charged under the 
nonperturbative $U(1)_X$ gauge field \cite{59}. 
 
We will illustrate how the anomaly arises on a 5-dimensional toy model, 
containing both 9-branes and 5-branes. 
Let us consider a 5-dimensional supergravity theory compactified on
${\bf R}^4 \times {\bf S}^1/Z_2$. The gauge degrees of freedom consist 
of  a 5-dimensional vector superfield $A_\mu$, {\em even under the 
 Ho\u{r}ava-Witten $Z_2$ parity} and a 4-dimensional vector
superfields $A_m^a$ which propagate on the boundary of spacetime. The
matter fields consist of: (i)   bulk hypermultiplets
$(\phi^a_{+}, \phi^a_{-}, \psi^a_{+}, \psi^a_{-})$, where the subscripts $+$ 
and $-$ denote the $Z_2$ parity of the corresponding field, $\phi_{\pm}$
are complex scalar fields and $\psi_{\pm}$ are Weyl fermions, (ii)
ordinary 4-dimensional matter living on the boundary of spacetime.

As is well-known \cite{GST}, the  supersymmetric Lagrangian describing
the interactions of the vector superfields is  determined by a
function ${\cal N}(\xi^C, \xi^A)$, a homogeneous cubic polynomial of
the coordinates $\xi^C$ and $\xi^A$, which are in correspondence with
the vector bosons $C^\mu$ and $A^\mu$. The scalar component of the
$U(1)_X$ vector multiplet parametrizes a  one-dimensional hypersurface 
of equation ${\cal N}(\xi^C, \xi^A) = 1$ in the 2-dimensional manifold of
coordinates $(\xi^C, \xi^A)$. These variables have an Horava-Witten
parity which is the opposite of the parity of the corresponding vector
field, that is $-1$ for $\xi^A$ and $+1$ for $\xi^C$.\footnote{ This is
more easily seen when compactifying to 4 dimensions where $\xi^C$
(resp. $\xi^A$) lies in the same supermultiplet as $C_5$ (resp. $A_5$).}
Therefore, the function ${\cal N}$ compatible with the Horava-Witten parity  
is simply: 
\begin{equation}
{\cal N}(\xi^C,\xi^A) =  (\xi^C)^3 
- {3 \over 2} \xi^C (\xi^A)^2  \ , 
\end{equation}
where we have chosen the normalisation in such a way that one recovers
diagonal and conveniently normalized gauge kinetic terms at the point $\xi^A
= 0$ (see below).

The corresponding Lagrangian for the
5-dimensional gauge fields reads:
\begin{eqnarray}
{\cal L} &=& {1 \over \pi} \int d^5 x \left[ \sqrt{g^{(5)}}
\left( -{1 \over 4} G_{_{AA}} F^{\mu \nu} F_{\mu\nu} 
-{1 \over 2} G_{_{AC}} F^{\mu \nu} C_{\mu\nu}  
-{1 \over 4} G_{_{CC}}C^{\mu \nu}  C_{\mu\nu} \right) \right. \nonumber \\ 
& & \; \; \; \left. +{1 \over 8} \epsilon^{\mu \nu \rho \sigma \lambda}
C_{\mu\nu} C_{\rho \sigma} C_\lambda
-{3 \over 16} \epsilon^{\mu \nu \rho \sigma \lambda}
F_{\mu\nu} F_{\rho \sigma} C_\lambda \right], \label{eq:gfLag} 
\end{eqnarray}
where $F_{\mu \nu} = \partial_\mu A_\nu - \partial_\nu A_\mu$ is our
$U(1)_X$ field strength,  $C_{\mu \nu} = \partial_\mu C_\nu - 
\partial_\nu C_\mu$ is the graviphoton field strength and
$G_{_{\Lambda\Sigma}} = -(1/2) \partial_{_{\Lambda}}\partial_{_{\Sigma}}
\ln {\cal N}|_{_{{\cal N}=1}}$. The topological
terms are fixed by the requirements of gauge invariance and supersymmetry 
in 5 dimensions. 

After compactification to 4 dimensions, one finds two chiral
supermultiplets of respective scalar fields:
\begin{eqnarray}
T^C &=& t \xi^C + i C_5 \ ,\nonumber \\
T^A &=& t \xi^A + iA_5 \ .
\end{eqnarray}
We work now in Planck mass units where $t=e^\gamma$. The normalisation
of the kinetic terms is evaluated at ${\cal N}(t\xi^C,t\xi^A)=t^3$
and, since $T^A$ does not survive the Ho\u{r}ava-Witten projection, at
$\xi^C = 1$. The K\"ahler potential for the remaining scalar
field $T^C$ is simply \cite{GST} $-\ln {\cal N}= -3 \ln (T^C + \bar
T^C)$. Taking into account the fact that the gauge field $C_m$ does not
survive either the  Ho\u{r}ava-Witten projection, the 4-dimensional
Lagrangian reads, including the scalar kinetic terms:
\begin{eqnarray}
{\cal L} & = & \int d^4 x \left( \sqrt{g} \left[ -{3 \over 8}t F^{mn}
F_{mn}+ {3 \over (T^C + \bar T^C)^2} \partial^m T^C \partial_m \bar T^C
\right] \right. \nonumber \\
& & \left. - {3 \over 8} {\rm Im} T^C F_{mn} \tilde F^{mn} \right)
 \ , \label{nonper}
\end{eqnarray}
Comparing the gauge kinetic term with (\ref{per0}), we see  the $S \leftrightarrow T$
exchange characterizing perturbative-nonperturbative mapping in 4 dimensions
\cite{5brane}. 

Our model  is
somewhat reminiscent of a model discussed recently by 
Mirabelli and Peskin \cite{MiPe} and we follow the method devised by
these authors to couple 4-dimensional boundary fields with
the fields living in the  5-dimensional bulk.

In the following we need the propagators of the bulk hypermultiplets, which we
compute in the limit of interest  $R_5 M_5 \gg 1$. 
 The Kaluza-Klein decomposition 
reads\footnote{ From now on, we use the dimensionless angular variable
$Mx_5$, where $M$ is the fundamental mass unit $m_{Pl} \sqrt{2\pi}$ 
introduced earlier, and
for simplicity still denote it by $x_5$ ($-\pi <x_5< \pi$).}:  
\begin{eqnarray}
(\phi_{+}, \psi_{+})= {1 \over \sqrt{2 \pi}} \sum_{n=0}^{\infty} 
\cos nx_5 \ (\phi_{+}^{(n)},\psi_{+}^{(n)}), \nonumber \\
\ (\phi_{-}, \psi_{-})= {1 \over \sqrt{2 \pi}} \sum_{n=1}^{\infty} \sin nx_5 \  
(\phi_{-}^{(n)} , \psi_{-}^{(n)}) \ . \label{p1} 
\end{eqnarray} 
The Feynman propagators are then computed to be
\begin{eqnarray}
<0|T \phi_{+}(x,x_5) \phi_{+}^{*}(y,y_5)|0> &=& {1 \over 2 \pi}
\sum_{n=-\infty}^{\infty} \cos nx_5 \cos ny_5 \ \Delta_F^{(n)} \ , \nonumber \\
<0|T \phi_{-}(x,x_5) \phi_{-}^{*}(y,y_5)|0> &=& {1 \over 2 \pi}
\sum_{n=-\infty}^{\infty} \sin nx_5 \sin ny_5 \ \Delta_F^{(n)} \ ,
\nonumber \\ 
<0|T \phi_{+}(x,x_5) \phi_{-}^{*}(y,y_5)|0> &=& 0 \ , \nonumber \\
<0|T \psi_{+}(x,x_5) \bar \psi_{+}(y,y_5)|0> &=& {1 \over 2 \pi}
\sum_{n=-\infty}^{\infty} \cos nx_5 \cos ny_5 \ S_{F,++}^{(n)} \ , \nonumber \\   
<0|T \psi_{-}(x,x_5) \bar \psi_{-}(y,y_5)|0> &=& {1 \over 2 \pi}
\sum_{n=-\infty}^{\infty} \sin nx_5 \sin ny_5 \ S_{F,--}^{(n)} \ , \nonumber \\
<0|T \psi_{+}(x,x_5) \psi_{-}(y,y_5)|0> &=& {1 \over 2 \pi}
\sum_{n=-\infty}^{\infty} \cos nx_5 \sin ny_5 \ S_{F,+-}^{(n)} \ ,
\label{p2}
\end{eqnarray}
where for example the fields $\phi^{(-n)}_{\pm}=\phi^{(n)}_{\pm}$, $\Delta_F^{(n)}$ is the Feynman
propagator for a complex scalar field of mass $m_{n}^2 = n^2/R_5^2$ and
we defined the massive fermion propagators
\begin{eqnarray}
S_{F,++}^{(n)}(x,y) &=& \int {d^4 p \over (2\pi)^4}
e^{-ip(x-y)}\;{ {\bar \sigma}^m p_m \over p^2
-m_n^2+i \epsilon} \ , \nonumber \\  
S_{F,+-}^{(n)}(x,y) &=& \int {d^4 p \over (2\pi)^4}
e^{-ip(x-y)}\;{m_n \over p^2 -m_n^2+i \epsilon} \ , \label{p3}
\end{eqnarray}  
and $S_{F,--}^{(n)}= S_{F,++}^{*(n)}$. By using a Schwinger proper-time 
representation, we can write, for example, 
\begin{equation}
<0|T \phi_{+}(x,x_5) \phi_{+}^{*}(y,y_5)|0>= {1 \over 2 \pi}
\int {d^4 p \over (2 \pi)^4} {e^{-ip(x-y)} \over p^2+i \epsilon} 
J(x_5,y_5, pR_5) \ , \label{p4}
\end{equation}
where
\begin{equation}
J(x_5,y_5, pR_5)=\int_0^{\infty} dt e^{-t} \sum_n \cos nx_5 \cos ny_5 
e^{-{tn^2 \over p^2R^2}} \simeq \pi 
[ \delta (x_5+y_5) + \delta (x_5-y_5) ] \ ,  \label{p5}
\end{equation}
where, in the last step, we have evaluated the function $J$ in the
ultraviolet region $p \sim M_5$, that is, in the region  of 
interest (\ref{p0}),  $pR_5 \gg 1$. Indeed, in the following,
we are interested in computing the triangle gauge anomaly and the
induced Fayet-Iliopoulos term, the computation of which involves 
the ultraviolet behaviour of the propagators.
In this case, the sum over Kaluza-Klein modes can be approximated 
by an integral. The final result is \footnote{All the $\delta$ functions
in our paper are periodic of period $2 \pi$, $\delta (x_5)= \delta (x_5-2 \pi)$.}:
\begin{eqnarray}
<0|T \phi_{+}(x,x_5) \phi_{+}^{*}(y,y_5)|0> &=& {1 \over 2}
[ \delta (x_5-y_5) + \delta (x_5+y_5) ] D_F(x-y) \ , \nonumber \\
<0|T \phi_{-}(x,x_5) \phi_{-}^{*}(y,y_5)|0> &=& {1 \over 2 }
[ \delta (x_5-y_5) - \delta (x_5+y_5) ] D_F (x-y) \ ,  \nonumber \\   
<0|T \phi_{+}(x,x_5) \phi_{-}^{*}(y,y_5)|0> &=& 0 \ , \\
<0|T \psi_{+}(x,x_5) \bar \psi_{+}(y,y_5)|0> &=& {1 \over 2 }
\  [ \delta (x_5-y_5) + \delta (x_5+y_5) ] S_F(x-y) \ , \nonumber \\   
<0|T \psi_{-}(x,x_5) \bar \psi_{-}(y,y_5)|0> &=& {1 \over 2 }
[ \delta (x_5-y_5) - \delta (x_5+y_5) ] S_F(x-y) \ , \nonumber \\
<0|T \psi_{+}(x,x_5) \psi_{-}(y,y_5)|0> &=& \frac{1}{2R_{5}}
[ \delta' (x_5-y_5) - \delta' (x_5+y_5) ] D_F(x-y), \ \nonumber
\label{p6}
\end{eqnarray}
where $D_F(x-y)$ and $S_F(x-y)$ are the Feynman propagators for a 
4-dimensional
complex massless boson and massless Weyl fermion, respectively.
Consistently with our previous discussion, we neglect $\delta'$ type
terms and therefore the last correlator in (\ref{p6}) is zero in the
limit (\ref{p0}). 

\begin{figure}[ht]
\centerline{\epsfxsize=\textwidth\epsfbox{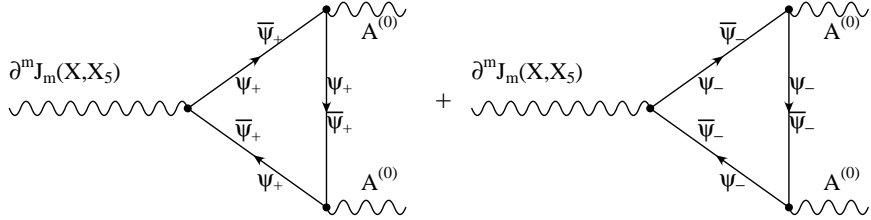}}
\caption{\it Triangle diagram for the $U(1)_X^3$ anomaly.}
\label{fig1}
\end{figure}

Using these results, we first compute the $U(1)_X^3$ triangle anomaly contribution
coming from bulk hypermultiplets containing the fermions $\psi_{+}^a,
\psi_{-}^a$ of charges $\pm X_{+}^a$, described by the 5-dimensional 
Dirac fermions
$\Psi^a=\left(\begin{array}{c}\psi^a_{+} \\ \bar\psi^a_{-}\end{array}
\right) $ (we use the Weyl basis
in the following) and from the boundary fermions $\psi_{\varphi ,i}$ living on the
two boundaries $i=1,2$. More precisely, 
we compute (see figure \ref{fig1})
$<~0| \partial^m J_m | \gamma^{(0)} \gamma^{(0)} |0>$, where $\gamma^{(0)}$
is the zero-mode of the $U(1)_X$ gauge field and  
\begin{eqnarray}
 J_m(x,x_5) &=& \sum_a X_{+}^{a}{\bar \Psi}^a (x,x_5) \gamma_m \Psi^a (x,x_5)
+ \sum_i \delta (x_5-x_{5,i}) X_{\varphi ,i} {\bar \psi_{\varphi,i}} {\bar \sigma_m} 
  \psi_{\varphi,i} \nonumber \\
&=& \sum_a \left[ \! X_{-}^{a}{\bar \psi}_{-}^a \! (x,x_5) \! \bar \sigma_m \! \psi_{-}^a (x,x_5) \!+\!
X_{+}^{a}{\bar \psi}_{+}^a (x,x_5) \! \bar \sigma_m \! \psi_{+}^a (x,x_5) \right]
\nonumber \\
&&+ \sum_i \! \delta (x_5 \!-\! x_{5,i}) \! X_{\varphi ,i}{\bar \psi_{\varphi ,i}} \! {\bar \sigma_m}
  \!  \psi_{\varphi ,i} 
\ , \label{p7}
\end{eqnarray}

where $X_{-}^{a}=-X_{+}^{a}$.   

In this computation, we are using the following identity, valid for periodic delta functions:
\begin{equation}
\delta (2x_{5})=\frac{1}{2}\left[\delta(x_{5})+\delta(x_{5}-\pi) \right]. \label{p70}
\end{equation}

By using the above propagators (\ref{p6}), we find
\begin{eqnarray}
&& \partial^m J_m (x,x_5) = \ A_{+} + A_{-} +\sum_i A_{\varphi ,i} =  \\
&& \left[ (Tr X_{\varphi ,1}^3 +  {1 \over 2} Tr X_{+}^3 )  
\delta (x_5)   +  
(Tr X_{\varphi ,2}^3  +  {1 \over 2} Tr X_{+}^3 )  \delta (x_5-\pi) \right]
 {g_4^2 \over 16 \pi^2} F^{mn} {\tilde F}_{mn} \ , \label{p8} \nonumber
\end{eqnarray}
where $A_{+}$, $A_{-}$ and $A_{\varphi,i}$ are the contributions of $\psi_{+}$, $\psi_{-}$
and respectively ${\psi}_{\varphi,i}$ running in the loop, as in figure
\ref{fig1}, and their
explicit expression read
\begin{eqnarray}
A_{+} \! &=& \! {1 \over 4} \left[ \delta (x_5) +\delta (x_5-\pi)+ 2 \delta (0) \right] 
Tr X_{+}^3 {g_4^2 \over 16 \pi^2} F^{mn}
{\tilde F}_{mn} \ , \nonumber \\
A_{-} \! &=& \! -{1 \over 4} \left[ \delta (x_5) + \delta (x_5-\pi)- 2 
\delta (0) \right] Tr X_{-}^3 {g_4^2 \over 16 \pi^2} F^{mn}
{\tilde F}_{mn} \ , \nonumber \\
A_{\varphi,i} &=&  \delta (x_5-x_{5,i}) Tr X_{\varphi,i}^3 {g_4^2 \over 16 \pi^2} F^{mn}
{\tilde F}_{mn} \ , \nonumber \\
\label{p80}
\end{eqnarray}
 
Strictly speaking, there are other diagrams contributing to the
anomaly, which however, give a result proportional to $\delta^{''}$ functions
and are therefore consistently neglected
within our hypothesis. A similar computation for $\partial^5 J_5$ gives
a similar result involving $\delta^{''}$ .
Similar computations can be made for the other, mixed gauge anomalies. The corresponding
anomaly coefficients are 
\begin{eqnarray}
&&U(1)_X^3 \ : \ \ \sum_i (Tr X_{\varphi,i}^3 + {1 \over 2} Tr X_{+}^3) \delta (x_5-x_{5,i})
\ , \nonumber \\
&&U(1)_X \ : \ \ \sum_i (Tr X_{\varphi,i} + {1 \over 2} Tr X_{+}) \delta (x_5-x_{5,i})
\ , \nonumber \\  
&&U(1)_X G_1^a G_1^b \ : \ \ \delta^{ab} \sum_{R_a} Tr X_{\varphi,1} T(R_a) \delta (x_5)
\ , \nonumber \\
&&U(1)_X G_2^c G_2^d \ : \ \ \delta^{cd} \sum_{R_c} Tr X_{\varphi,2} T(R_c) \delta (x_5-\pi)
\ , \nonumber \\
&&U(1)_X G_1^a G_2^c \ : \ \ 0 \ . \label{p83}
\end{eqnarray}
In (\ref{p83}), $G_i$ stand for gauge groups on the boundary $i$ and $T(R)$ for Dynkin
index of charged fermions. The last mixed anomaly is automatically zero as no fermion
living on the boundaries can be simultaneously charged under $G_1$ and $G_2$ gauge groups.
The anomaly-free conditions are easily read from (\ref{p83}) 
which, if violated, signal the presence of a anomalous $U(1)_X$ in the theory.

It is interesting to notice the close analogy between (\ref{p8}) and the modified
Bianchi identity \cite{HoWi} (written here in differential form language)
\begin{equation}
d G = -{3 \sqrt 2 \over 2 \pi} \left( \frac{\kappa_{11}}{4\pi} \right)^{2 \over 3} \sum_i \delta (x_5-x_{5,i}) (tr F_i^2-{1 \over 2}
tr R^2) dx^5 \ , \label{p81}
\end{equation}
where $G$ is the field strength of the three form , $F_i$ is the gauge field form on
the boundary $i$ and $trR^2$ is computed from the curvature two form. 
The interpretation of (\ref{p8}) is, of course, similar to (\ref{p81}), i.e.
the anomaly coming from the 5d hypermultiplets is equally distributed on the two
boundaries, which was expected physically. 
If we integrate over $x_5$ in (\ref{p8}) in order to find the global anomaly term,
we find, as usual
\begin{equation}
\partial^m J_m^{(0)} = (Tr X_{\varphi ,1}^3 + Tr X_{\varphi ,2}^3 + 
 Tr X_{+}^3 ) {g_4^2 \over 16 \pi^2} F^{mn} {\tilde F}_{mn} \ . \label{p82}
\end{equation}

\begin{figure}[ht]
\centerline{\epsfxsize=\textwidth\epsfbox{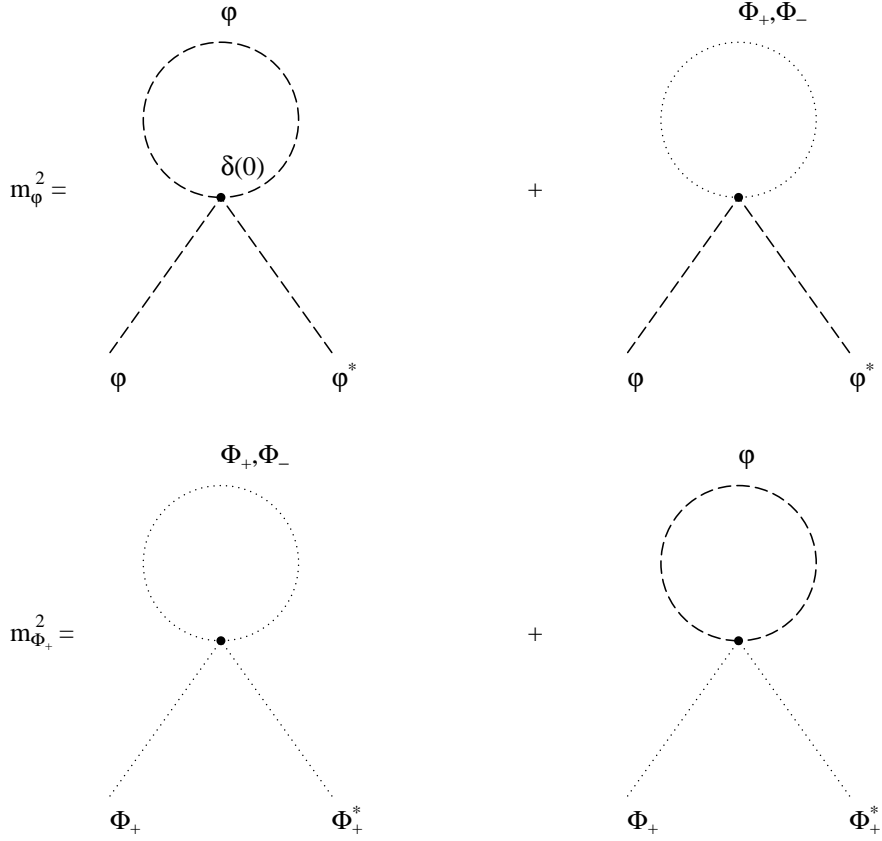}}
\caption{\it One loop diagrams involved in the computation of the induced mass terms for the
charged scalar fields, on the boundary ($\varphi$) or in the bulk ($\Phi$).}
\label{fig2}
\end{figure}

We now come to our main goal, the computation of the one-loop 
Fayet-Iliopoulos terms induced through the orbifold $Z_2$ projection.
To accomodate the general case, we assume that both the fields
living on the boundaries (denoted by
$\varphi ,i$ in the following) and the bulk fields contribute to the
anomaly and therefore generate Fayet-Iliopoulos terms.
These can be simply found by computing the induced mass terms for the
charged scalar fields, of either boundary or bulk type, to be obtained
from the diagrams shown in figure \ref{fig2}. The results, obtained by using the
interaction terms:

\begin{eqnarray}
V_D &=& {g_5^2 \over 2} \int d^5 x \sqrt g \left[ X_{+} \phi_{+} 
{\partial K \over \partial \phi_{+}} + X_{-} \phi_{-} {\partial K \over
\partial \phi_{-}}   \nonumber \right.\\
  & & \left. + \sum_i \delta (x_5-x_{5,i}) X_{\varphi ,i} \varphi_i {\partial K
\over \partial \varphi_i} \right]^{2}
\end{eqnarray}

 and the propagators (\ref{p6}) are
\begin{eqnarray}
m_{\varphi ,i}^2= g_4^2 \delta (0) (Tr X_{\varphi ,i}+{1 \over 2} Tr X_{+})
X_{\varphi ,i} {1 \over 192 \pi^2} M_5^2 \ , \nonumber \\
m_{X_+}^2= g_4^2  (Tr X_{\varphi,1}+Tr X_{\varphi,2}+
Tr X_{+}) X_{+} {1 \over 192 \pi^2} M_5^2 \ , \label{p9} 
\end{eqnarray}

where we choose the ultraviolet regulator in figure \ref{fig2}, in complete analogy
with the weakly-coupled case, as in the second reference in \cite{DSW}.
More precisely, we cut-off the momentum integration at $p^2=(1/3)M_5^2$,
as  the effective field theory description breaks down above
$M_5^2$ and the numerical factor $1/3$ is due to the stringy cut-off 
\cite{DSW}.

The results can be interpreted as the generation in the effective lagrangian
of a Fayet-Iliopoulos term on each boundary $\xi_i$, such that the $U(1)_X$ D-term in the
scalar potential reads
\begin{eqnarray}
V_D &=& {g_5^2 \over 2} \int d^5 x \sqrt g \left[ X_{+} \phi_{+} 
{\partial K \over \partial \phi_{+}} + X_{-} \phi_{-} {\partial K \over
\partial \phi_{-}}   \nonumber \right.\\
  & & \left. + \sum_i \delta (x_5-x_{5,i}) X_{\varphi ,i} \varphi_i {\partial K
\over \partial \varphi_i} + \sum_i \delta (x_5-x_{5,i}) \xi_i \right]^2 \ , 
\label{p10}
\end{eqnarray}

where 
\begin{equation}
\xi_1 = (Tr X_{\varphi ,1}+{1 \over 2} Tr X_{+}) {1 \over 192 \pi^2} M_5^2 \ , \\
\xi_2 = (Tr X_{\varphi ,2}+{1 \over 2} Tr X_{+}) {1 \over 192 \pi^2} M_5^2 \
\label{p11}
\end{equation}
and $g_4^2 = {2 \pi g_5^2 / t}$.
In analogy with (\ref{p8}), here also the global Fayet-Iliopoulos term, obtained by
integrating over $x_5$ the above densities is the expression  
\begin{equation}
\xi = (Tr X_{\varphi ,1}+Tr X_{\varphi ,2}+ Tr X_{+}) {1 \over 192 \pi^2} M_5^2 \ , 
\label{p12}
\end{equation} 
very similar to the one obtained in the perturbative heterotic string where 
in (\ref{p12}) $M_5$ is replaced by $m_{Pl}$.

This gives a slightly lower scale of breaking for the anomalous
$U(1)_X$, which goes in the right direction for solving the cosmological
issues related with such a symmetry \cite{Dterm,cosmo}.

A supersymmetry preserving vacuum in (\ref{p10}) is found if on each boundary there is
at least one scalar field $\varphi ,i$ of charge opposite in sign to the corresponding
Fayet-Iliopoulos term $\xi_i$, which takes a compensating v.e.v. \footnote{A v.e.v. for a bulk 
field of order $ \delta(x_{5}-x_{5,i})\xi_i$ would not minimize the kinetic energy density.}  
 
\begin{equation}
\varphi_1=-{\xi_1 \over X_{\varphi,1}} \ , \varphi_2=-{\xi_2 \over X_{\varphi,2}}
\ . \label{p13}
\end{equation}
  
Supersymmetry can be broken in this context if a gaugino condensate $<\lambda \lambda >$
forms on one
boundary ($2$, for concreteness, with a Fayet-Iliopoulos term $\xi_2$) and the
corresponding F-term condition is incompatible with the $U(1)_X$ D-term condition. As in
\cite{BiDu}, the resulting soft terms in the observable sector (boundary $1$) are
\begin{equation}
{\tilde m} \sim { < \lambda \lambda > \over \xi_2 } \ . \label{p130}
\end{equation}   
The usual supergravity-induced soft terms by the gaugino condensation \cite{NOY} are 
here much
smaller, therefore the anomalous $U(1)_X$ contributions are dominant.
Phenomenologically relevant soft terms ask therefore for a condensate scale 
$< \lambda \lambda > \sim (10^{11} GeV)^3$, lower than the analogous one in the
perturbative heterotic case.

The generalization of the above results to the case of more than one
perturbative $C^{\alpha}_{\mu}$ and nonperturbative gauge fields
$A^i_{\mu}$ is straightforward. The five-dimensional bosonic spectrum 
related to gauge fields contains  the gravitational multiplet 
$(g_{\mu \nu}, C_{\mu})$, perturbative vector multiplets
$(C_{\mu}^{\alpha}, \xi_C^{\alpha})$, $\alpha = 1 \cdots
h_{1,1}-1$, odd under $Z_2$ projection and even nonperturbative vector multiplets $(A_{\mu}^i , \xi_A^i )$, where $h_{1,1}$
characterizes the 6-dimensional compact manifold with the intersection 
numbers $c_{\alpha \beta \gamma}$. The 5-dimensional prepotential 
describing the vector action is
\begin{equation}
{\cal N} = {1 \over 6} c_{\alpha \beta \gamma} \xi_C^{\alpha} \xi_C^{\beta}
\xi_C^{\gamma} - {1 \over 2} c_{\alpha ii} \xi_C^{\alpha} ( \xi_A^i )^2
\ . \label{p14}
\end{equation}
The four-dimensional action is found by putting ${\cal N}=1$,
$\xi_A^i=0$ and the 4-dimensional
even complex moduli are
\begin{equation}
T_C^{\alpha} = t \xi_C^{\alpha} + i C_5^{\alpha} \ . \label{p15}
\end{equation}
The 4d gauge kinetic function of the nonperturbative gauge fields then read
\begin{equation}
f_{ij} = -{1 \over 2} \partial_i \partial_j \ln {\cal N} |_{{\cal N}=1, \xi_A^i=0}
= c_{\alpha ij} T_C^{\alpha}
\end{equation}
and are therefore non-universal, depending on compactification
details. The 
4-dimensional Green-Schwarz mechanism will generalize accordingly, 
involving more
moduli axions able to shift gauge anomalies, in analogy with the 
six-dimensional case \cite{S}.

\vskip .5cm
{\bf Note added}

After this work was completed, we received the preprint \cite{JMR}, where the question
of anomalous $U(1)_X$ in M-theory and open strings was studied. Our arguments
seem to indicate that the universal anomalous $U(1)_X$ of \cite{DSW} considered in \cite{JMR} cannot 
give a one-loop Fayet-Iliopoulos term in M-theory 
and rather describes some unkonwn, 
nonperturbative physics. 
     



\end{document}